# DataHub: Collaborative Data Science & Dataset Version Management at Scale

Anant Bhardwaj[1], Souvik Bhattacherjee[2], Amit Chavan[2]
Amol Deshpande[2], Aaron J. Elmore[1], Samuel Madden[1], Aditya Parameswaran[1,3]

[1]MIT, [2]U. Maryland (UMD), [3]U. Illinois (UIUC)
anantb@mit.edu, bsouvik|amitc|amol@cs.umd.edu, aelmore|madden@csail.mit.edu, adityagp@illinois.edu

## ABSTRACT

Relational databases have limited support for *data collaboration*, where teams collaboratively curate and analyze large datasets. Inspired by software version control systems like `git`, we propose (a) a dataset version control system, giving users the ability to create, branch, merge, difference and search large, divergent collections of datasets, and (b) a platform, DATAHUB, that gives users the ability to perform collaborative data analysis building on this version control system. We outline the challenges in providing dataset version control at scale.

## 1. INTRODUCTION

The rise of the Internet, smart phones, and wireless sensors has produced a huge diversity of datasets about all aspects of our lives, from our social interactions to our vital signs and medical records to events and happenings across the globe. There has also been a proliferation of large datasets thanks to the open data initiatives adopted by numerous governments and organizations, and increasing effectiveness of scientific instruments at collecting data like genomic data, astronomical data, etc. These datasets are quite varied in nature, ranging from small to very large, from structured (tabular) to unstructured, and from largely complete to noisy and incomplete; further, the datasets typically evolve over time, often at very rapid rates. Increasingly, researchers and "data science" teams want to collect, analyze, and collaborate on these datasets, to extract actionable insights or to distill scientific knowledge from it. Such collaborative data analysis is often ad hoc, featuring significant back-and-forth among the members of the team, and also trial-and-error to identify the right analysis tools, programs, and parameters. Such collaborative analysis fundamentally requires the ability to keep track of, and reason about, the datasets being used. In essence, what is needed is a system to track *dataset versions* over time: the operations performed on these datasets by multiple individuals; the new datasets that originate as a result; and other external data products that depend on these datasets. At the same time, to reduce the barrier for entry, there is a need for a distributed hosted platform that simplifies and facilitates the collaborative analysis process by providing reliable "always on" access to different versions of data.

Current data management systems are inadequate at supporting these types of collaborative environments. Relational databases work well when data conforms to a schema, but often these datasets consist of a mix of relational and less structured (or differently structured) data. Furthermore, relational systems (as we describe below) lack support for version management. Using version control systems, such as `git`, for large datasets is also known to have several shortcomings [22]. This leads to teams resorting storing data in file systems, often using highly ad hoc and manual version management and sharing techniques. It is not uncommon to see directories containing thousands of files with names like *data1-v1.csv, data1-v2.csv, data1-v1-after-applying-program-XYZ.txt,* etc., possibly distributed and duplicated across multiple cloud storage platforms. Thus, while there is a wealth of tools for "doing" data science, no previous tools have explicitly addressed the problem of dataset versioning, especially when many users are analyzing, collaborating, modifying, and sharing datasets.

To fill this gap, we propose two tightly-integrated systems. First, Dataset Version Control System (DSVC), is a system for multiversion dataset management. *DSVC's goal is to provide a common substrate to enable data scientists to capture their modifications, minimize storage costs, use a declarative language to reason about versions, identify differences between versions, and share datasets with other scientists.* Second, DATAHUB, is a hosted platform built on top of DSVC, that not only supports richer interaction capabilities, but also provides a number of novel tools for data cleaning, data search and integration, and data visualization tools. Here we draw an analogy to the popular source code version control solutions, `git` and `github`—DSVC is similar to `git` but supports a significantly richer query language and scales to larger and more structured datasets, whereas DATAHUB is analogous to `github`, but also provides more complete features for working with structured data.

To illustrate the need for DSVC, we briefly describe results of an informal survey of how data collaboration works in a several computational biology groups at MIT. We believe the example is representative of many other data science teams in a variety of domains. Key properties of these data collaborations are as follows:

- Teams of 20-30 students, postdocs, faculty, and researchers, share approximately 100 TB of data via a networked file system.
- This storage costs about $800 / TB / year, for maintenance costs from a local storage provider for unlimited read/write access (the same storage on Amazon S3 would cost about $400 / TB / year, excluding access costs.) This amounts to nearly $100K/year, which is a substantial cost for an academic research group.
- When a new researcher wants to work with data, he or she may make a private copy of the data, or may read it from a shared folder. Researchers have no way of knowing which other researcher's programs refer to their shared data folders.
- There are significant but unknown amounts of duplication of data. Simple file-level duplicate detection is insufficient since there is often some modification or extension in duplicates.
- Space (or cost) constraints lead to frequent requests from the PI-in-charge to others to reduce storage. This leads to removal of intermediate versions and data products, and causes stress as researchers don't know who is using their data or whether a particular dataset is essential for reproducibility of some experiment. If the dataset could be deterministically re-generated, or if it were possible to track when a dataset had last been used by whom, researchers would be more comfortable with deleting these datasets.
- Researchers would prefer a transparent mechanism to access data and write versions in order to be backward compatible with pre-existing scripts (which are coded using a file-level API.)
- Researchers do not make heavy use of metadata management tools (like relational databases, wikis, etc) due to perceived cost of adding and maintenance of that metadata. There have been several failed wiki attempts.

This scenario makes it clear that an effective dataset version management system could substantially reduce costs by eliminating redundant storage, help users find and manage relevant data and the relationships between data items, and help them share and exchange both datasets and derived products and collaboratively analyze them. Similar dataset versioning scenarios arise in many sub-communities. For example, consider a group of sociologists wrangling and cleaning datasets from the web, ecologists integrating a set of biodiversity studies with varying measurement techniques, a city that wants to publish transit ridership data for mash-ups, or a team of medical data analysts and doctors that need to clean and annotate ECG data for building models on seizures.

While there has been some work on dataset versioning, e.g., [2, 18, 14, 20, 19, 16, 17], this work has been focused on managing and querying a linear chain of versions, such as would arise when a single user is modifying their own data. In contrast, in the applications above, different users often branch off from a common version, and frequently need to merge their branches back together to perform consolidated analysis or publish a new version of a data set. Furthermore, a common task involves comparing different versions of data sets, e.g., to study the relative effectiveness of two algorithms, treatments, analyses, etc.

Our proposed DSVC system aims at providing such rich version management functionality, for both structured or unstructured datasets. Building DSVC requires addressing many novel research challenges, most of which have not been addressed in prior work.

- DSVC needs to be able to handle large datasets (100s of MBs to 100s of GBs), a large number of files per dataset, and a large number of versions. Existing version control systems (VCS) do not scale along either dimension (Table 1), because (somewhat surprisingly given their popularity) they employ fairly rudimentary techniques underneath [22]. To handle the scale that we envision, we need to develop novel and highly efficient algorithms for identifying and estimating differences between versions, for choosing storage layouts in a workload-aware fashion, for deciding which versions to store fully and which in a compressed form, and for answering different types of queries.
- The basic querying and retrieval functionality supported by existing VCS is inadequate for the purposes of scientists or data analysts, who would like to declaratively query specific versions of datasets (using say SQL), explore and query provenance information across datasets (e.g., using a provenance query language [9]), analyze differences between versions, and identify version(s) that satisfy certain properties. Designing a unified query language that supports these diverse classes of queries is a major research challenge that we need to address; further, to our knowledge, the last two types of queries have not been systematically studied in prior work.
- Existing VCS, primarily aimed at maintaining source code, view data as uninterpretable collections of bytes. However, unlike source code, most datasets exhibit structure that can be explicitly represented along with the data. DSVC needs to support in situ declarative querying on such datasets, which is a ma-jor challenge given that most datasets are not stored in their entirety, i.e., some dataset versions may only be stored as modifications from other versions of the same datasets. DSVC also needs to support analysis, exploration, and provenance queries across versions (e.g., finding when a record was last modified, or determining the user who has inserted the largest number of records). This requires tracking modifications and differences between versions at the record level. At the same time, we need to develop new techniques that can efficiently exploit the structure to reduce the space required to store versions.

| Method | Load Time | Data Size | Read One Version |
|---|---|---|---|
| Unencoded | 574.5 s | 16.0 GB | 192.0 s |
| Compressed | 2,340.4 s | 2.01 GB | 18.63 s |
| SVN | 8,070.0 s | 16.0 GB | 29.2 s |
| Git | – | – | – |

**Table 1: SVN and Git Performance on Map Data**

- Finally, existing VCS have limited support for trigger-like functionality, or *hooks*, where actions are taken in response to changes to the repository. To handle rapidly evolving or streaming data, such functionality needs to be made a first-class citizen, to enable automated analysis, to keep derived data products up-to-date, and to identify major changes in the data (e.g., a data source changing the schema it uses to provide data). How to specify triggers and how to efficiently execute a large number of triggers written using the rich query language that we discussed above are major research challenges that need to be addressed.

As noted above, we are building DATAHUB, a hosted version control system built around DSVC. DATAHUB provides a range of additional functionality aimed at facilitating and simplifying data analysis built upon a multi-versioning infrastructure, including access control for groups and individuals, support for annotating versions with text and structured metadata, and support for data cleaning, version search, integration, differencing, and visualization tools.

In the rest of this paper, we focus on the version management features in DSVC, in particular on the versioning API and query language (§3), and the challenges associated with building efficient implementations of these interfaces within DATAHUB (§4).

## 2. PRIOR WORK

There has been a long line of source code version control systems, from CVS to SVN to git. These systems, however, are designed to deal with modest-sized files and do not scale well to large numbers of large files. To show this, we measured the time to load a collection of 16 versions of a 1 GB road network database into `git` and `svn` and the time to read a single version (we ran `svnadmin pack` and `git repack` after after loading all data to compress it). With 8 GB of RAM, git ran out of memory (it compares all versions against each other in an inefficient manner that uses very large amounts of RAM for large files), and svn was very slow to encode differences, taking about 20x longer than the time to load and copy an uncompressed raw file. In addition, as discussed in the previous section, these systems provide rudimentary querying or analytics capabilities. We ran the experiment on an Intel Xeon quad-core (Intel E7310), with 8GB of RAM and a single 7200RPM SATA drive running Ubuntu Lucid.

Most data science systems, including analytic packages like SAS, Excel, R, Matlab, Mahout, Scikit (`scikit-learn.org`), and Pandas (`pandas.pydata.org`); workflow tools like Pegasus [4], Chimera [6], and VisTrails [1]; and collaboration tools like Fusion tables [12], Orchestra [8], and CQMS [10, 7] lack dataset versioning management capabilities. We envision DSVC could serve as the backend data management layer for many of these systems.

There are also several recent startups and projects on providing basic dataset management infrastructure for data science applications, including CKAN (ckan.org), Domo (domo.org), Enterprise Data Hub (cloudera.com/enterprise), Domino (dominoup.com), Amazon Zocalo and Dat Data (dat-data.com). The emergence of these systems shows that the capabilities that we aim to provide are sorely needed by practitioners; at the same time, to our knowledge, none of these tools provide the rich versioning and querying functionality we describe in this paper.

Versioning has been studied recently in the context of specific types of data like graphs [11] and arrays [17]: these papers identify preliminary techniques for differencing these dataset types and performing queries over the versions. We plan to build on that work, focusing on more general structured datasets, and on developing compact methods for encoding and querying the differences between datasets. Besides this prior work there has been relatively little work on versioning in database literature. One exception is Buneman et al. [2], which proposed a linear versioning system for collections of hierarchical documents, comparing delta-chain and periodic materialization approaches. It was not, however, a full-fledged version control system representing an arbitrary graph of versions; instead focusing on algorithms for compactly encoding a linear chain of versions. Temporal databases [18, 14, 20, 19, 16] and Oracle's "Flashback" feature, provide, in effect, the ability to traverse and query a linear chain of versions, but lack support for branching versions or derived data products. Additionally, some scientific workflow systems and provenance systems provide the ability to manage and traverse arbitrary derivation (provenance) graphs [3], but lack version control features.

## 3. PROPOSED DESIGN AND INTERFACES

In this section, we present a high level overview of the various components of DSVC and DATAHUB, the interfaces they expose, and applications built on top or as part of them.

### 3.1 High Level Design

The high level architecture of DATAHUB and DSVC is depicted in Figure 1. The core of DSVC is the DSVCP: the dataset version control processor, which processes and manages versions. The DSVCP exposes the versioning API (described in §3.3) for client applications to use. At a high level, the versioning API is similar to the git API, but comes with additional functionality. The version query processor (VQP) is tightly integrated with the DSVCP, and exposes VQL (described in §3.5) as an interface to client applications. In short, VQL is an enhanced version of SQL that allows users and applications to query multiple versions at once.

DATAHUB is a layer on top of VQP and DSVCP, and consists of everything within the dashed lines. DATAHUB is a hosted platform, designed to be run as a server that many clients use to store their data; there will both be a public DATAHUB site, as well as potentially many private DATAHUBs run by organizations. DATAHUB provides the versioning API and the VQL to applications. In addition to these two interfaces, DATAHUB also provides a variety of basic functionality aimed at reducing the burden to data analytics, including data ingest and integrate (also shown in the diagram).

In addition to the interfaces and data manipulation functionality, DATAHUB includes fine-grained access control features to allow datasets to be declared public/private and read-only/writeable by different sets of users. We plan to adopt standard techniques for role-based access control [5]. In addition, we will provide language bindings to make possible to access DATAHUB datasets from popular languages, including Javascript, enabling the sharing of web pages that directly access and manipulate persistent data.

We have implemented an early prototype of this system that uses PostgreSQL as the backend, creating a separate database for each user and providing basic per-user access control. It exposes a Thrift-based [21] API that wraps queries and query results in Thrift objects (Thrift is a cross platform data marshaling layer designed to compactly exchange data in a platform-independent manner). This makes it possible to invoke DATAHUB queries and retrieve results from any of the 20+ languages supported by Thrift. We have also built a simple object-relational mapping layer for Python and Java, allowing programmers to access DATAHUB datasets as though they are persistent objects in these languages. Our prototype includes some initial versioning features, described below. Subsequently, we plan to replace this simple backend with our custom storage layer that supports efficient versioning, provenance tracking, visualization, and other features, with the goal of keeping a constantly running prototype up and running (our current simple prototype is running at http://datahub.csail.mit.edu).

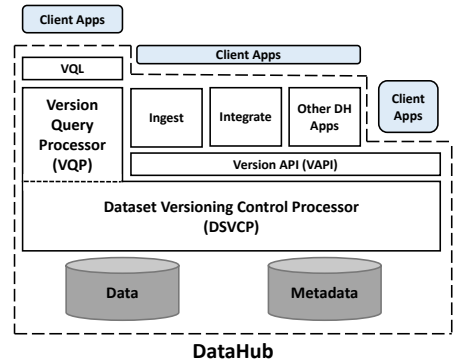

Figure 1: DataHub Components and Architecture.

### 3.2 DSVC Data Model

We now describe the underlying data model that DSVC espouses. At a high level, DSVC uses a flexible "schema-later" data representation, that allows data to be represented in all forms, from textual blobs, to key-value pairs, to fully structured records. (The reason for this is that we expect users to store carefully curated structured data as well raw data in various stages of cleaning and integration.)

The main abstraction that DSVC uses is that of a *table* containing *records*. We adopt a simple model where every record in a table has a *key*, and an arbitrary number of typed, named attributes associated with it; we use the word *schema* to loosely refer to the attributes associated with a record. We expect that there will be large groups of records with identical schemas in a table, with a fully curated table having the same schema for every record. For completely unstructured data, a single key can be used to refer to an entire file. While there are many other models for data that relax relational "schema-first" requirements, we chose this model because it offers a flexible approach for storing a wide range of data at different levels of structure, allowing both unstructured and fully structured data to co-exist. We do not intend this choice to be limiting or controversial; other "semi-structured" representations, e.g., XML or JSON, would suffice as well.

The second abstraction that DSVC uses is that of a *dataset*. A dataset consists of a set of tables, along with any correlation / connection structure between them (e.g., foreign key constraints).

The versioning information is maintained at the level of datasets in the form of a *version graph*. Version graph is a directed acyclic graph where the nodes correspond to the datasets and the edges capture relationships between versions as well as the *provenance* information that records user-provided and automatically inferred annotations for derivation relationships between data. A directed edge from node $u$ to node $v$ indicates that either $v$ is a new version of $u$, or $v$ is a new *branch* that is created as a copy of $u$ and will evolve separately (so long as it is not merged back in), or $v$ is a derived data product obtained by applying an operation to $u$. Whenever a version of a dataset is created in DATAHUB, users or applications may additionally supply *provenance metadata* that indicates the relationship between two versions. This metadata is designed allow queries over the relationship between versions, and includes both *fine-grained* and *coarse-grained* provenance data. Different applications may store different provenance data, but examples include (i) the name of the program that generated the new version, (ii) the commit id of the program in a code version control system like git (if available), (iii) the identifiers of any other datasets that may have been used to create the new version. In addition, we may store the list of records in the original version that contributed to a changed record in the new version if that information is available and compactly representable.

## 3.3 Versioning API

DSVC provides a versioning API similar to what version control systems like git provide, including the following commands:
- `create`: create a new dataset
- `branch`: create a new version of a dataset; future updates to this branch will not be reflected in the original dataset.
- `merge`: merge two or more branches of the dataset
- `commit`: make local (uncommitted) changes to the dataset permanent. Can be used in a fine-grained (transactional) way or to merge batches updates. In particular, like git, many private transactions can be run by a user, and can be pushed as a batch.
- `rollback`: revoke changes to the dataset
- `checkout`: create a local copy of branch that is either a full copy, a lazily retrieved copy, or a sampled partial copy (see §3.4).

In addition to the commands above, the versioning API, like git, allows users to specify `hooks`. Similar to triggers, `hooks` are used to run scripts before, during, and after commits to code. For example (adapted from githooks.com), checking commit messages for spelling errors, enforcing coding standards, letting people know of a new commit, or moving code to production. In our case, we expect `hooks` to encompass all the typical use-cases, such as enforcing standards or continuous integration, along with additional cases, such as notifying applications to run off the newest copy dataset (e.g., a dashboard plotting aggregate statistics), or tracking data products derived from the dataset (e.g., via aggregation or filtering). In these cases, `hooks` may be used to rerun the application or update the derived data products respectively. `hooks` could also be used in lieu of triggers to detect and correct errors.

This basic API can be used by end users and DATAHUB components such as data cleaning and integration tools. For example, a user who uploads genomic data from a spreadsheet and then does cleaning and integration with other datasets would first create the new dataset, propagate and commit raw data to it, and then perform several branching steps as cleaning and integration are completed.

## 3.4 Checking Out, Branching, and Merging

Distributed version control systems enable users to easily create and retrieve copies of a dataset. A user can branch (or fork) a dataset to make isolated changes that may be merged back into another version at a later time.

Users of DATAHUB can interact with datasets in two ways: either by directly issuing queries to DATAHUB servers ("transactional mode"), or by checking out local copies of datasets which they manipulate and commit changes to ("local mode"). Transactional mode works just as in a traditional relational system: concurrent transactions are isolated from each other using locking, and the system provides a guarantee of serial equivalence.

| Query Input | | Query Outputs | |
|---|---|---|---|
| Data & Version | SQL on version | VQL (find similar versions) | |
| Data | SQL on master | VQL (find version matching constraint) | |
| | Data | | Version |

**Figure 2: Sample DataHub Queries.**

Local mode requires more complicated machinery to deal with conflicting updates, since two users may both modify the same version of the same records in their local dataset. We allow the users control over how much of the dataset to copy out at once, and how much to fetch as needed "lazily", so as to make working with large datasets substantially more efficient in local mode.

In addition to checking out local versions of a dataset, DATAHUB will enable users to checkout a sampled version of a dataset. With a sampled dataset, users will perform updates on the local sampled copy and when the user merges the sampled copy back to the repository, DATAHUB will apply the modifications to the entire dataset and notify the user of any constraint violations or merge conflicts. We plan to limit certain operations when working on a sample to ensure that a user is aware of the missing records. An example of a limited operation is preventing updates to specific rows; instead we require updates based on predicates so that DATAHUB can apply the update to the entire dataset.

DSVC allows users to create a branch of a dataset either in local or transactional mode. It is critical that branching is is a quick process that minimizes the amount of data required for each branch. Sec. 4 highlights the research challenges associated with creating versions of large datasets in DSVC.

**Research Challenges:** The utility of creating multiple branches of a dataset increases with the ability to merge divergent branches. Otherwise, users end up with many variations of a dataset with an unknown number of dependencies. In building DSVC we will explore how to define a conflict, how to easily detect conflicts between branches, how to merge non-conflicting divergent branches, and how to guide a user through merging conflicting branches. Traditional source control software define conflicts by concurrent modifications to the same line. This is the semantic equivalent of row-level conflicts for structured datasets. However, two branches may alter disjoint attributes for overlapping rows. Therefore in this scenario a user may choose to apply the modifications from both branches as there would be no lost update. We will leverage existing research in disconnected databases, commutative and replicated data structures, and operational transformations for detecting and resolving conflicts in simultaneous/collaborative editing systems. In building DATAHUB we plan to explore various definitions of dataset branch conflicts, using deltas to detect conflicts, techniques to summarise large differences between branches, and tools to aid users in resolving conflicts that arise from a merge.

## 3.5 Versioning Query Language

In addition to the API, DSVC supports a powerful versioning query language based on SQL, called VQL, that allows users to operate on one or more versions at a time, returning results that are either data items or pointers to versions. Conceptually, a VQL query could specify whether it should be executed only on the tables that contains all the fields specified in the query (the current default), or on tables containing some but not all of the fields. We describe our preliminary design for VQL via examples below.

Consider the chart in Figure 2, which displays four ways in which VQL can be used. The square in the lower left corner is the most straightforward way to use VQL: standard SQL queries that will be executed on the master version (by default). Since the query may contain one or more predicates, this way the query input involves "data", and the output is once again "data". On the other hand, the square in the upper left corner allows us to specify which version or versions we would like the standard SQL queries to be executed. For instance, VQL supports the query

```
SELECT * FROM R(v124), R(v135)
WHERE R(v124).id = R(v135).id
```

where v124, v135 are version numbers. Once again, the query specifies "data", but also specifies one or more "versions".

The squares in the right hand side are a bit different: in this case, the result is one or more version numbers. Here, we add to SQL two new keywords: VNUM and VERSIONS, which can be used in the following manner:

```
SELECT VNUM FROM VERSIONS(R)
WHERE EXISTS (SELECT * FROM R(VNUM)
WHERE name = 'Hector')
```

This query selects all versions where a tuple with name Hector exists. The attribute VNUM refers to a version number, while VERSIONS(R) refers to a special single-column table containing all the version numbers of R. The example above is a VQL query that fits in the right bottom corner of the chart, while a VQL query that provides a version as input and asks for similar versions (based on user-specified predicates) would fit into the right top corner.

```
SELECT VNUM FROM VERSIONS(R)
WHERE 10 > DIFF_RECS(R, VNUM, 10)
```

where DIFF_RECS is a special function that returns the number of records that are different across the two versions. VQL will support several such functions that operate on versions (e.g., DISTANCE(R, 10, 20) will return the derivation distance between the versions 10 and 20 of R (the result is -1 if 20 is not a descendant of 10 in the version graph).)

Naturally, there are examples that span multiple regions in the quadrant as well: as an example, the following query selects the contents of a relation $S$ from the first time when a large number of records were added between two versions of another relation $R$ in the same dataset.

```
SELECT * FROM S(SELECT MIN(VR1.VNUM) FROM
VERSIONS(R) VR1, VERSIONS(R) VR2
WHERE DISTANCE(R,VR1.VNUM,VR2.VNUM)=1
AND DIFF_RECS(R,VR1.VNUM,VR2.VNUM)>100)
```

**Research Challenges:** The query above is somewhat unwieldy; fleshing out VQL into a more complete, easy-to-use language is one of the major research challenges we plan to address during our work. In particular, we would like our eventual query language to be able to support the following features, as well as those discussed above, while still being usable:

- Once a collection of VNUMs is retrieved, performing operations on the data contained in the corresponding versions is not easily expressible via VQL as described. For example, users may want the ability to use a for clause, e.g., do X for all versions satisfying some property. For this, concepts from nested relational databases [15] may be useful, but would need further investigation.
- Specifying and querying for a subgraph of versions is also not easy using VQL described thus far; for this, we may want to use a restricted subset of graph query languages or semi-structured query languages.
- Users should be able to seamlessly query provenance metadata about versions, as well as derived products (specified via hooks), in addition to the versions, e.g., find all datasets that used a specific input tuple found to be erroneous later, or find datasets that were generated by applying a specific cleaning program.

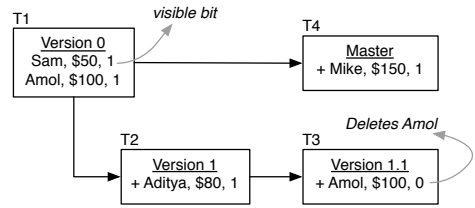

**Figure 3: Example of relational tables created to encode 4 versions, with deletion bits.**

In addition to VQL, which is a SQL-like language, DSVC will also support a collection of flexible operators for record splitting and string manipulation, including regex functionality, similarity search, and other operations to support the data cleaning engine, as well as arbitrary user-defined functions.

## 4. STORAGE REPRESENTATIONS

In this section, we describe two possible ways to represent a version graph: the *version-first* representation, where, for each version, we (logically) store the collection of records that are a part of that version, possibly in terms of deltas from a chain of parent versions. The second way of representing dataset versions is what we call a *record-first representation*, where we (logically) store each record, and for each record, we store the (compressed) list of versions that that record appears in. We describe these two approaches in turn.

### 4.1 Version-First Representation

The version-first representation is the most natural, because, as in git-like systems, it makes it easy for users to "check out" all of the records in a particular version.

Abstractly, we can think of encoding a branching history of versions in a *storage graph*, with one or more fully materialized versions, and a collection of deltas representing non-materialized versions. Retrieval queries can be answered by "walking" this storage graph appropriately. Note that nodes in this storage graph may not have a one-to-one correspondance with nodes in the version graph, as we may want to add additional nodes to make retrieval more efficient. We describe this idea in more detail below.

For relational datasets, it is relatively straightforward to emulate this abstract model in SQL. Whenever the user performs a branch command, we simply create a new table to represent changes made to the database after this branch was created. This new table has the same schema as the base table. In addition, each record is extended with a *deleted* bit that allows us to track whether the record is active in a particular version. To read the data as a particular version, a we can take the union of all of the ancestor tables of a particular version, being careful to filter out records removed in later versions. In addition, updates need to be encoded as deletes and re-insertions. An example of this approach is shown in Figure 3. Here, there are two branches. At the head of the "Master" branch, the table contains Sam, Amol, Mike. At the head of the Version 1 branch (labeled "Version 1.1"), the table contains Sam, Aditya because the Amol has been marked as deleted. It is possible to implement this scheme completely in SQL, in any existing database using simply filters and union queries. Of course, the performance may be suboptimal, as lots of UNIONs and small tables can inhibit scan and index performance, so investigating schemes that encode versions below the SQL interface will be important. Additionally, non-relational datasets may be difficult to encode in this representation, requiring other storage models.

In the rest of this section, we describe challenges in implementing this version-first representation, in either the SQL-based or inside-the-database setting.

**Challenge 1: Recording Deltas:** Given two versions, how can we record the delta between them compactly so that we can retrieve one version using the other version and the delta? Note here that the two versions may correspond to different tables and may have different schemas. In the simple SQL implementation described above, versions are created explicitly via INSERT/DELETE commands. However, we also plan to support the creation of versions in external tools as well. For differencing such arbitrary versions the simplest approach would be to use a binary differencing algorithm, such as "diff" or "bdiff". Such tools work by finding common substrings in files. Unfortunately, they are memory intensive and can be slow, requiring many seconds to compare large files.

We plan to investigate database-aware approaches that express deltas in terms of the specific records that were inserted/deleted, or in terms of portions of records that are modified. Finding such differences efficiently, when the query or program that transformed the two versions is unknown, is challenging. We will explore database-log like structures to encode the modifications from one database to another, coupled with search algorithms to efficiently find differences. Making such techniques work on large datasets (that exceed the memory size) will require algorithms to quickly find physical blocks that differ (e.g., using hash-trees [13]), and then searching around the regions of difference to identify record-level deltas that "explain" the physical differences that were identified.

A related challenge is also to be able to *estimate* the *size* of the difference between two versions quickly. This is crucial for deciding which versions to delta against each other as we discuss below.

**Challenge 2: Version Graph Encoding:** The problem of efficiently encoding a graph of versions is also challenging. Just because two versions are adjacent in the version graph doesn't mean that they should be stored as differences against each other. For example, for the version graph depicted in Figure 3, depending on the exact changes made, the difference between T3 and T4 may be smaller than the difference between T3 and T2.

There is also a trade-off between materializing a specific version or just storing a delta from a past version, since following delta chains may be expensive. We plan to employ optimization techniques to find the optimal encoding of the version graph by considering all possible pairwise encodings. Further, we plan to consider adding what we call *Steiner* datasets to optimize retrieval—analogous to Steiner points, these "extra" datasets can be used to reduce the retrieval costs as demonstrated in our work on historical graph data management [11]. Effective heuristics will be needed because this search space is very large.

### 4.2 Record-first Representation

In this section we describe the record-first representation, where we encode data as a list of records, each annotated with the versions it belongs to. The advantage of this representation over the version-first representation is that it makes it easy to find all versions that records with particular properties participate in, as we can use an index to find satisfying records, and then retrieve their versions. It also lends itself to storage representations that are clustered by record properties, rather than being clustered according to the containing version.

A simple example of a record-first representation is a *temporal database*, which is essentially a linear chain of versions, where each record is annotated with an interval that denotes the "versions" it belongs to. For a branching array of versions, we would need to explicitly encode the versions that contain a record using a (compressed) bitmap – a temporal interval is essentially equivalent to a run-length-encoding of such a bitmap.

From a querying perspective, certain types of queries are more efficient to execute against such a representation. Consider the query that asks for all versions that satisfy a certain property: this can be executed efficiently by taking an OR of the bitmaps of all satisfying records. Similarly if we want to execute a particular query against a group of versions (e.g., an aggregate query), this storage representation naturally enables sharing of computation. On the other hand, a retrieval query (i.e., fetch a given version) may end up doing redundant work since the union is likely to be much larger than any specific version.

### 4.3 Storage vs. Efficiency

In describing the two representations above, we have alluded to a fundamental tradeoff that permeates all aspects of dataset version management: the tradeoff between *efficiency* and *storage*. Simply put, since versions share a lot of information, we can store versions very compactly, but retrieving and running queries on them may become expensive. While the efficiency-versus-storage tradeoff certainly exists for databases without versioning as well, the tradeoff is exacerbated here because many versions are similar to each other, and storing them in a naive fashion could lead to severe issues with efficiency or with storage, as described in the introduction.

**Prioritizing Efficiency:** While it is certainly possible to just use one representation, for DSVC to be efficient at both API operations as well as VQL queries, both representations are needed. At a high level, retrieving complete versions is much more efficient in the version-first representation, while identifying versions that have a certain property is much more efficient in the record-first representation. On the other hand, retrieving versions that satisfy some property may work best by combining representations: the record-first to identify the appropriate versions, and the version-first to retrieve the requisite versions. In addition to using multiple representations, it is certainly possible to also use indexing and caching of prior VQL query results to quickly identify or recreate often-queried versions.

**Prioritizing Storage:** Finally, periodic archival and cleaning up of versions is essential to reduce storage costs and keep storage within the user-configurable budget. Deciding how to incrementally meet the budget constraint by deleting or storing in a compressed form (via deltas from existing versions) infrequently accessed versions will be a challenge. We expect to use `hooks` to "clean up" after such an operation.

## 5. CONCLUSIONS

In this paper, we introduced two tightly-integrated systems: a dataset version control system, and DATAHUB, drawing an analogy with `git` and `github`. We outlined the many challenges in managing and querying large multi-version datasets that do not apply to regular source-code version control, and proposed initial solutions. We believe addressing these challenges is essential in supporting large-scale collaborative data analytics.